\documentclass[%
 twocolumn, 
superscriptaddress,
preprintnumbers,
 amsmath,amssymb,
 aps, 
pra,
floatfix,
]{revtex4-2}
\usepackage{graphicx}
\usepackage{dcolumn}
\usepackage{bm}
\usepackage{amsmath}
\usepackage{braket}
\usepackage{color}
\usepackage{comment}
\usepackage{siunitx}
\usepackage[normalem]{ulem}

\begin{document}
\newcommand{\Jnature}{Nature (London)}
\newcommand{\Jnatphys}{Nat. Phys.}
\newcommand{\Jnatcomm}{Nat. Comm.}
\newcommand{\Jnatmat}{Nat. Mater.}
\newcommand{\JSciRep}{Sci. Rep.}

\newcommand{\Jscience}{Science}
\newcommand{\Jsciadv}{Science Advances}

\newcommand{\Jpnas}{Proc. Nat.l Acad.  Sci.}

\newcommand{\Jprx}{Phys. Rev. X}
\newcommand{\Jprl}{Phys. Rev. Lett.}
\newcommand{\Jpr}{Phys. Rev.}
\newcommand{\Jpra}{Phys. Rev. A}
\newcommand{\Jprb}{Phys. Rev. B}
\newcommand{\Jprc}{Phys. Rev. C}
\newcommand{\Jprd}{Phys. Rev. D}
\newcommand{\Jpre}{Phys. Rev. E}
\newcommand{\Jrmp}{Rev. Mod. Phys.}

\newcommand{\JplA}{Phys. Lett. A}

\newcommand{\Jepl}{Europhys. Lett.}
\newcommand{\Jnjp}{New J. Phys.}
\newcommand{\Jepjb}{Eur. Phys. J. B}
\newcommand{\Jepjd}{Eur. Phys. J. D}
\newcommand{\Jepjst}{Eur. Phys. J. Special Topics}

\newcommand{\crasphy}{C. R. Phys.}

\newcommand{\Jjosab}{J. Opt. Soc. Am. B}
\newcommand{\JApplPhysLett}{Appl. Phys. Lett.}
\newcommand{\JApplPhysB}{Appl. Phys. B}

\newcommand{\Joptcomm}{Opt. Comm.}
\newcommand{\JApplOpt}{Appl. Opt.}

\newcommand{\JphysA}{J. Phys. A}

\newcommand{\JphyslettA}{J. Phys. Lett. A}
\newcommand{\JphyslettB}{J. Phys. Lett. B}

\newcommand{\JnulcphysA}{Nucl. Phys. A}

\newcommand{\Jbullamphyssoc}{Bull. Am. Phys. Soc.}

\newcommand{\Jcpl}{Chin. Phys. Lett.}

\newcommand{\Jphysjap}{J. Phys. Soc. Japan}

\newcommand{\JJlowT}{J. Low Temp. Phys.}

\newcommand{\Jprocroysoc}{Proc. Roy. Soc. A: Math. Phys. Eng. Sci.}

\newcommand{\Jjetp}{Sov. Phys. JETP}
\newcommand{\Jjetplett}{JETP Lett.}
\newcommand{\JjphysUSSR}{J. Phys. USSR}
\newcommand{\JSovJlowT}{Sov. J. Low Temp. Phys.}

\newcommand{\JZhEkspTeorFiz}{Zh. Eksp. Teor. Fiz.}

\newcommand{\Jjchemphys}{J. Chem. Phys.}
\newcommand{\Jjphyschemsol}{J. Phys. Chem. Sol.}
\newcommand{\Jsolstatecomm}{Solid State Comm.}

\newcommand{\Jijmpb}{Int. J. Mod. Phys. B}

\newcommand{\Jijthphys}{Int. J. Theor. Phys.}

\newcommand{\JphysicaB}{Physica B}
\newcommand{\JphysicaBC}{Physica B+C}

\newcommand{\Jstatmech}{J. Stat. Mech.}
\newcommand{\Jstatphys}{J. Stat. Phys.}

\newcommand{\Jphysrep}{Phys. Rep.}
\newcommand{\JRepProgPhys}{Rep. Prog. Phys.}
\newcommand{\JjphysCM}{J. Phys.: Cond. Matt.}
\newcommand{\JjphysA}{J. Phys. A: Math. Theor.}
\newcommand{\JjphysB}{J. Phys. B: At. Mol. Opt. Phys.}
\newcommand{\JjphysC}{J. Phys. C: Solid State Phys.}
\newcommand{\JjphysF}{J. Phys. F: Metal Phys.}
\newcommand{\Jphystoday}{Phys. Today}

\newcommand{\Jadvphys}{Adv. Phys.}

\newcommand{\Jannphys}{Ann. Phys. (NY)}

\newcommand{\JAnnualRevCondMat}{Annual Rev. Cond. Mat. Phys.}

\newcommand{\Jadvatmoloptphys}{Adv. At. Mol. Opt. Phys.}
\newcommand{\Joptexpr}{Opt. Express}

\newcommand{\JphysB}{J. Phys. B: At. Mol. Opt. Phys.}

\newcommand{\JTheorMathPhys}{Theor. Math. Phys.}
\newcommand{\JMathPhys}{J. Math. Phys.}
\newcommand{\JCommMathPhys}{Comm. Math. Phys.}

\newcommand{\Jprogthphys}{Prog. Theor. Phys.}
\newcommand{\Jprogthphyssup}{Prog. Theor. Phys., Suppl.}

\newcommand{\JjphysFr}{J. Phys. (France)}
\newcommand{\JjphysquatreF}{J. Phys. IV (France)}

\newcommand{\Jcommmathphys}{Comm. Math. Phys.}

\newcommand{\JZphys}{Z. Phys.}
\newcommand{\JZphysB}{Z. Phys. B}

\newcommand{\JRevSciInstrum}{Rev. Sci. Instrum.}

\newcommand{\JFortschrPhys}{Fortschr. Phys.}

\newcommand{\Jcompphyscom}{Comput. Phys. Commun.}
\newcommand{\Jphysconfser}{J. Phys. Conf. Ser.}

\newcommand{\Jphysicascripta}{Phys. Scr.}

\newcommand{\Jphysics}{Physics}

\newcommand{\JprocphyssocA}{Proc. Phys. Soc. A}

\title{\textbf{The wave nature of a Mott insulator} 
}%

\author{ Xudong Yu } \thanks{These authors contributed equally to this work.}
\affiliation{Institut f{\"u}r Experimentalphysik und Zentrum f{\"u}r Quantenphysik, Universit{\"a}t Innsbruck, Technikerstra{\ss}e 25, Innsbruck, 6020, Austria} 

\author{ Chengyang Wu } \thanks{These authors contributed equally to this work.}
\affiliation{Institute of Quantum Electronics, School of Electronics, Peking University, Beijing 100871, China} 

\author{ Wenhan Chen } 
\affiliation{Institut f{\"u}r Experimentalphysik und Zentrum f{\"u}r Quantenphysik, Universit{\"a}t Innsbruck, Technikerstra{\ss}e 25, Innsbruck, 6020, Austria} 

\author{ Igor Zhuravlev }
\affiliation{Institut f{\"u}r Experimentalphysik und Zentrum f{\"u}r Quantenphysik, Universit{\"a}t Innsbruck, Technikerstra{\ss}e 25, Innsbruck, 6020, Austria} 

\author{ Zekui Wang} 
\affiliation{Institut f{\"u}r Experimentalphysik und Zentrum f{\"u}r Quantenphysik, Universit{\"a}t Innsbruck, Technikerstra{\ss}e 25, Innsbruck, 6020, Austria} 
\affiliation{State Key Laboratory of Quantum Optics Technologies and Devices, Institute of Opto-electronics, Shanxi University, Taiyuan, Shanxi
030006, China}

\author{ Yi Zeng }
\affiliation{Institut f{\"u}r Experimentalphysik und Zentrum f{\"u}r Quantenphysik, Universit{\"a}t Innsbruck, Technikerstra{\ss}e 25, Innsbruck, 6020, Austria} 

\author{ Sudipta Dhar}
\affiliation{Institut f{\"u}r Experimentalphysik und Zentrum f{\"u}r Quantenphysik, Universit{\"a}t Innsbruck, Technikerstra{\ss}e 25, Innsbruck, 6020, Austria}

\author{ Milena Horvath}
\affiliation{Institut f{\"u}r Experimentalphysik und Zentrum f{\"u}r Quantenphysik, Universit{\"a}t Innsbruck, Technikerstra{\ss}e 25, Innsbruck, 6020, Austria}

\author{ Thierry Giamarchi}
\affiliation{DQMP, University of Geneva, 24 Quai Ernest-Ansermet, Geneva, CH-1211, Switzerland}

\author{ Manuele  Landini }
\affiliation{Institut f{\"u}r Experimentalphysik und Zentrum f{\"u}r Quantenphysik, Universit{\"a}t Innsbruck, Technikerstra{\ss}e 25, Innsbruck, 6020, Austria}

\author{ Hanns-Christoph  N{\"a}gerl}\email{christoph.naegerl@uibk.ac.at}
\affiliation{Institut f{\"u}r Experimentalphysik und Zentrum f{\"u}r Quantenphysik, Universit{\"a}t Innsbruck, Technikerstra{\ss}e 25, Innsbruck, 6020, Austria}

\author{ Hepeng Yao }\email{hepeng.yao@pku.edu.cn}
\affiliation{Institute of Quantum Electronics, School of Electronics, Peking University, Beijing 100871, China} 

\author{ Yanliang  Guo }\email{yanliang.guo@uibk.ac.at}
\affiliation{Institut f{\"u}r Experimentalphysik und Zentrum f{\"u}r Quantenphysik, Universit{\"a}t Innsbruck, Technikerstra{\ss}e 25, Innsbruck, 6020, Austria}
\affiliation{Key Laboratory of Quantum State Construction and Manipulation (Ministry of Education), School of Physics, Renmin University of China, Beijing 100872, China}

\date{\today}

\date{\today}

\begin{abstract}
Quantum phases of matter are routinely identified by coherence features, with interference patterns being one of the most directly observable quantities~\cite {andrews1997observation,anderson1966special,leggett2001bose}. In lattices, the superfluid-to-Mott-insulator (SF-MI) transition is commonly viewed as a change from wave-like coherence to particle-like localization: interference peaks are taken as a hallmark of superfluidity, whereas their disappearance is used to diagnose insulating behavior~\cite{greiner2002,bloch2008many}. Here, we challenge this picture for one-dimensional (1D) strongly interacting gases subject to a lattice potential. We realize a gapped Mott insulator through pinning in a shallow lattice and find that pronounced interference peaks persist deep in the insulating regime. Strikingly, the interference becomes stronger as the Mott fraction increases, demonstrating that 
a certain degree of coherence still exists in the insulator state.
Measurements of the one-body correlation function reveal an oscillatory, exponentially decaying coherence pattern across several lattice sites, in quantitative agreement with quantum Monte Carlo (QMC) simulations. Our work shows that interference does not uniquely diagnose superfluidity and it exposes the unexpected wave nature of a 1D Mott insulator.

\end{abstract}

\maketitle

Strongly correlated quantum matter supports a remarkable variety of phases, from dissipationless superfluids and incompressible Mott insulators (MI)~\cite{mott1949basis,mott2004metal,jaksch1998,greiner2002} to disordered Bose glasses~\cite{yao-boseglass-2020,Smith_NP_2016,gautier-2Dquasicrystal-2021,yu2024observing} and supersolids with intertwined order~\cite{Boninsegni2012, tanzi2019, bottcher2019, chomaz2019}. What distinguishes these states is not their constituent particles, but the many-body organization of coherence, fluctuations, rigidity and localization~\cite{sachdev2001}. A central aim of modern physics is to understand how such distinct forms of quantum order emerge and transform, especially when interactions drive the system across phase boundaries that reorganize its correlations and excitations~\cite{Sondhi1997}. Among the most basic of these are transitions between coherent and insulating phases, which probes how phase information survives, reshapes or disappears in the presence of strong correlations~\cite{bloch2008many}. 

Interacting lattice bosons provide a paradigmatic setting for this question through the SF-MI transition~\cite{stoferle2004transition,lewenstein2007ultracold,giamarchi1997mott}. The standard picture of this transition is generally built around interference: side peaks in the momentum distribution are widely interpreted as evidence of superfluidity, whereas their disappearance is taken to signal the onset of insulating behavior~\cite{greiner2002,bloch2008many}. This view mirrors a broader physical intuition. A superfluid is naturally associated with a coherent matter wave extending across the lattice, whereas a MI is commonly represented as localized particles with well-defined site occupation~\cite{gemelke2009,bakr2010}. In second-quantized language, this contrast is reflected by a coherent superfluid state, $(L^{-1/2}\sum_j \hat{a}_j^\dagger)^N|0\rangle$, versus the ideal Mott state $\prod_i \hat{a}_i^\dagger|0\rangle$ at unit filling~\cite{bloch2008many}. Interference has therefore become both a practical diagnostic and a conceptual dividing line between the two phases. Yet 1D quantum-liquid theory already suggests that this distinction need not be exact: In Haldane's harmonic-fluid description, a subleading oscillatory contribution to the one-body correlation function can persist even in the Mott-insulating regime and, in principle, generate interference upon expansion~\cite{haldane1981effective,giamarchi2003quantum,RevModPhys.83.1405}. However, since this contribution is much weaker than the leading term~\cite{RevModPhys.83.1405}, it poses a challenge for experimental verification. Whether a MI can retain a directly observable wave character has remained an open question.


\begin{figure*}[t]
\centering
\includegraphics[width=1\textwidth]{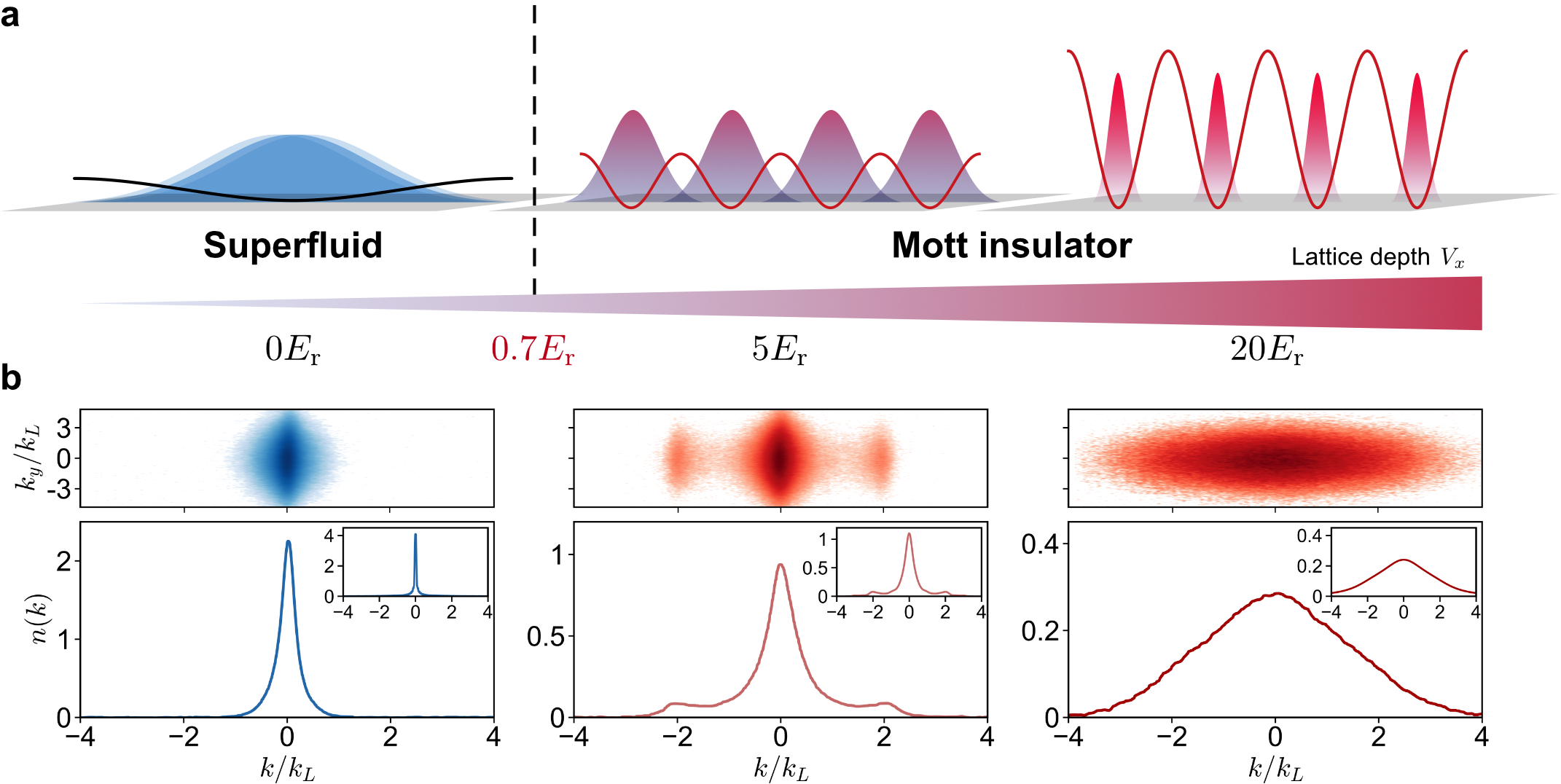}
\caption{\label{fig:1} 
\textbf{Real-space illustration and representative momentum distributions across the SF-MI transition.}
\textbf{a,} Schematic evolution of the longitudinal density in a strongly interacting 1D lattice gas as the lattice depth is increased, covering the transition from the SF ($V_x\!=\!0$) to the MI regime ($V_x\!=\!5E_\mathrm{r}$, $20E_\mathrm{r}$). \textbf{b,} Representative two-dimensional ToF images at $V_x\!=\!0$, $5$, and $20E_\mathrm{r}$ (top panel) and corresponding longitudinal $n(k)$ (bottom panel). Insets: QMC-calculated $n(k)$ for the same lattice depth and interaction strength as in the experiment but for a homogeneous 1D system in a box trap at near-zero temperature. All experimental data are averaged over 20 repetitions.
}
\end{figure*}

In this work, we report the observation of interference patterns for a MI, revealing its wave nature. By loading strongly-interacting 1D bosons into a shallow optical lattice, we prepare the MI in the pinning regime \cite{haller2010pinning,boeris2016mott}. We observe pronounced side peaks in the momentum distribution that persist deep in the gapped MI regime, and find that the interference pattern becomes more prominent as the Mott fraction increases, indicating that it arises from the insulating state itself rather than from residual superfluid components. Consistently, our QMC simulation results in the near-zero-temperature box-trap limit also exhibit interference for a nearly pure Mott state. To quantitatively uncover its microscopic origin, we determine the one-body correlation function and identify an oscillatory coherence pattern superimposed on exponential decay across several lattice sites. The corresponding QMC simulations, taking into account finite temperature and harmonic confinement, show quantitatively good agreement with our measurements. Together, these results establish that the 1D MI retains a directly observable wave character. They provide experimental access to the subleading second-harmonic correlations anticipated in Haldane’s harmonic-fluid description \cite{haldane1981effective,RevModPhys.83.1405}.


For the experiment, we start with a Bose-Einstein condensate (BEC) of approximately $1.2\times10^5$ $^{133}\mathrm{Cs}$ atoms, levitated and confined in a dipole trap \cite{Kraemer2004}. We initially prepare a near-unity-filled 3D MI state by adiabatically loading the BEC into a 3D optical lattice with depth $(V_x,V_y,V_z)\!=\!(20,30,30)E_\mathrm{r}$, with $x$ denoting the longitudinal axis and $y$ and $z$ the transverse axes, setting the s-wave scattering length to $a_s\!=\!300a_0$. Here, $a_0$ is Bohr's radius and $E_\mathrm{r}\!=\!\pi^2\hbar^2/(2ma^2)$ denotes the photon recoil energy with lattice spacing $a\!=\!\pi/k_\text{L}\!=\!532.2 \,\textrm{nm}$ and wave number $k_\text{L}$. We then adiabatically tune $a_s$ to the desired value via a Feshbach resonance. By decreasing $V_x$ to zero while keeping $V_y$ and $V_z$ at the initial large values, we realize a 2D array of independent 1D tubes for which the dynamics happens primarily along the $x$-direction, as illustrated in Fig.~\ref{fig:1}a. After a $20$-ms hold time at the target lattice depth, the optical potentials and the interaction are switched off and the atoms are detected by absorption imaging after $50$-ms time-of-flight (ToF). This allows the in-trap phase coherence to manifest itself as an interference pattern in the expanded cloud. After integrating over the transverse momentum components, we obtain the longitudinal momentum distribution $n(k)$, see Fig.\ref{fig:1}b.




\begin{figure*}[t]
\centering
\includegraphics[width=1\textwidth]{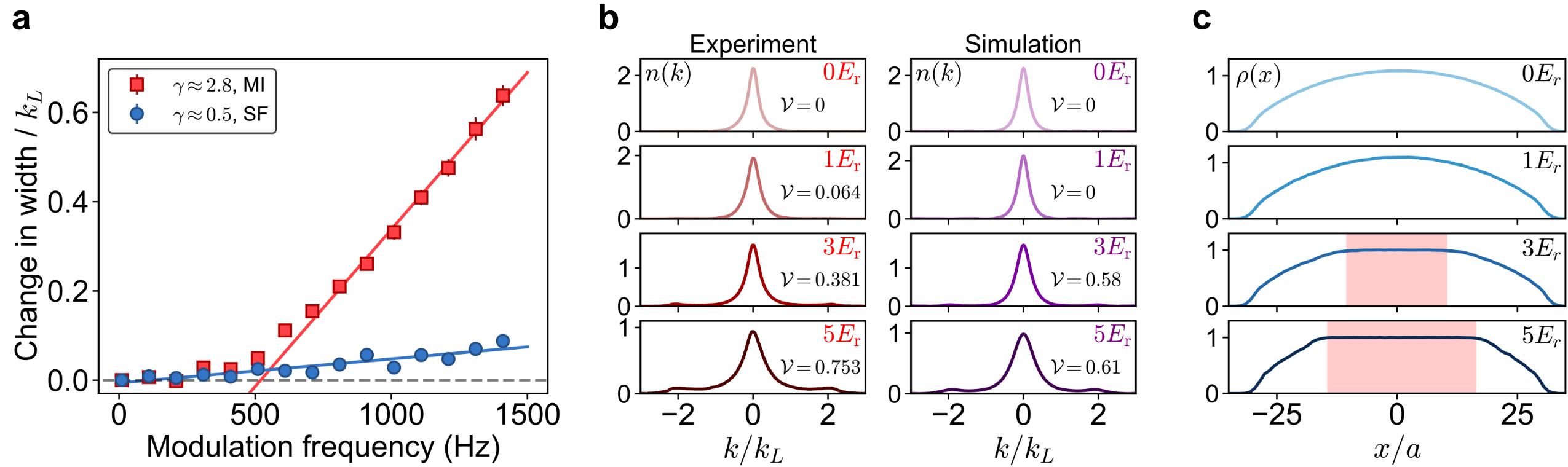}
\caption{\label{fig:2} \textbf{Mott gap and interference enhancement with increasing Mott fraction.}
\textbf{a,} Lattice-modulation spectroscopy at $V_x\!=\!1.5E_\mathrm{r}$. The increase in post-ToF cloud size after modulation is plotted as a function of modulation frequency $f$ for weak interactions ($\gamma\!=\!0.5$, blue circles) and strong interactions ($\gamma\!=\!2.8$, red squares). The solid lines indicate linear fits and the resulting gap for $\gamma\!=\!2.8$ is \SI{0.52(8)}{\kilo\hertz}.
\textbf{b,} Longitudinal $n(k)$ at $V_x\!=\!0$, $1$, $3$, and $5E_\mathrm{r}$ for experiments (left column) and QMC calculations (right column) including the harmonic trap. The visibility $\mathcal{V}$ of the $\pm2\hbar k_{\rm L}$ peaks is indicated in each panel.
\textbf{c,} Corresponding real-space density profiles from QMC calculations with harmonic confinement at the same lattice depths as in \textbf{b}. The blue regions denote the compressible superfluid component in the trap, whereas the shaded red area marks the incompressible MI fraction.
}

\end{figure*}

We set $a_s\!=\!220a_0$ to investigate the pinning SF-MI transition, giving the dimensionless Lieb-Liniger interaction parameter~\cite{giamarchi2003quantum,RevModPhys.83.1405} the value $\gamma\!\approx\!2.8$, for which the transition is expected near $0.7E_\mathrm{r}$~\cite{boeris2016mott,haller2010pinning}.
We then vary $V_x$ from $0$ to $20E_\mathrm{r}$, thereby traversing the transition from the superfluid to the insulating regime. At $V_x\!=\!0$, the distribution $n(k)$ exhibits a single narrow peak centered at zero momentum as shown in Fig.\ref{fig:1}b, consistent with a spatially extended and phase-coherent superfluid. For the case of a deep lattice $V_x\!=\!20E_\mathrm{r}$, the distribution $n(k)$ broadens into a nearly Gaussian profile, as expected for a well localized Mott state in which the atoms are tightly confined to the individual lattice sites \cite{jaksch1998,sherson2010single,greiner2002}. 
Strikingly, at an intermediate depth $V_x\!=\!5E_\mathrm{r}$, where the system is already placed inside the MI regime, the distribution $n(k)$ shows not only a narrow central peak but also clear side peaks at $k\!=\!\pm 2k_\mathrm{L}$. Such an interference pattern is usually regarded as a hallmark of superfluid phase coherence. Its presence in the insulating regime therefore reveals a coexistence of pinned particle motion and wave-like interference structure.

\begin{figure*}[t]
\centering
\includegraphics[width=0.98\textwidth]{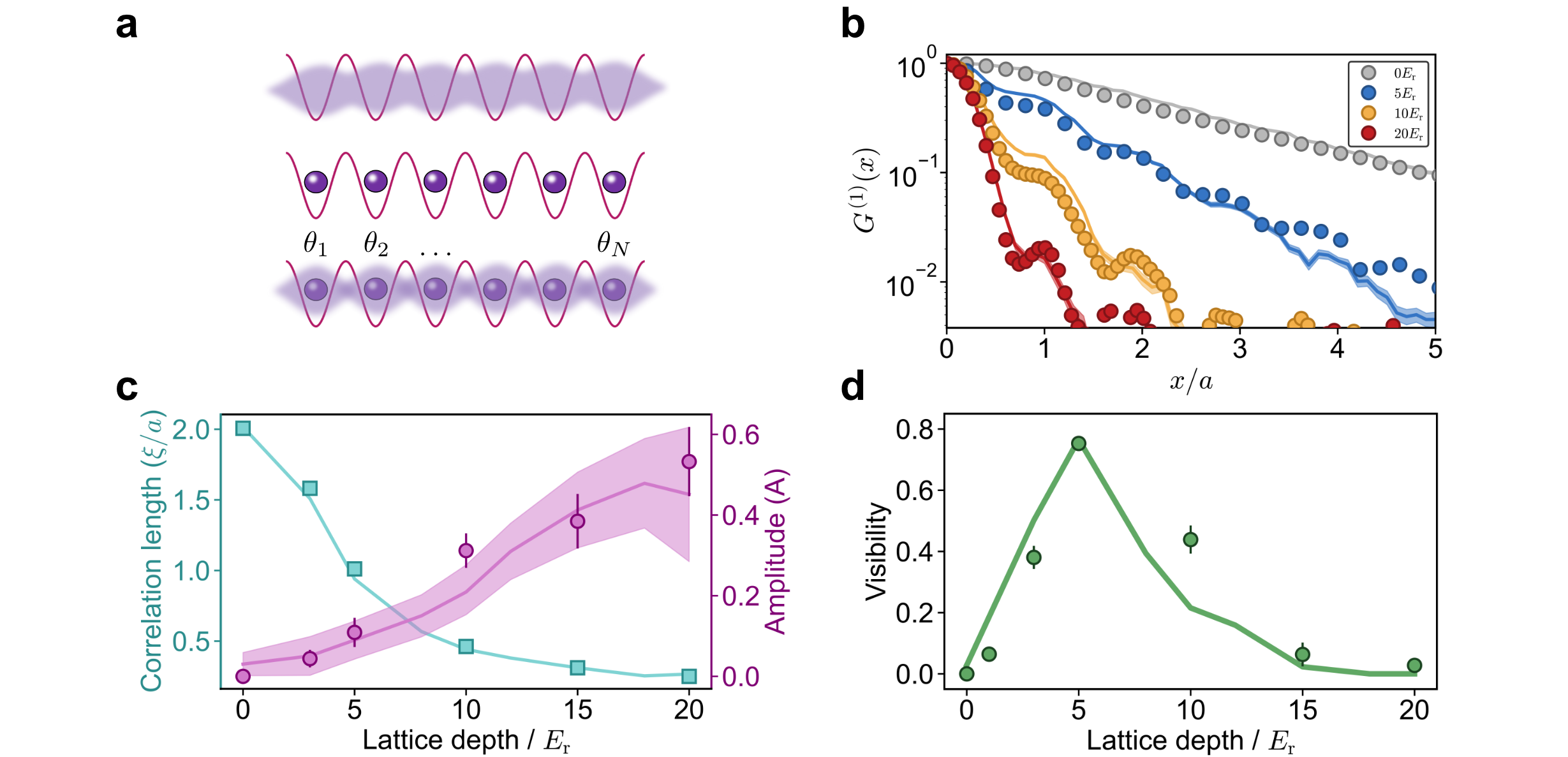}
\caption{\label{fig:4} \textbf{ Real-space picture and correlations underlying the interference.}
\textbf{a,} Conventional picture of the SF-MI transition where the superfluid is viewed as a coherent matter wave (top panel), while the MI is represented as particles localized on individual lattice sites (middle panel). Revised picture for the 1D MI with strong interactions in a shallow lattice (bottom panel), in which neighboring-site wavefunctions can still overlap, allowing phase structure and interference to persist, with $\theta_j$ denoting the local phase associated with site $j$.
\textbf{b,} One-body correlation function $G^{(1)}(x)$ at $V_x\!=\!0$ (grey), $5$ (blue), $10$ (gold), and $20E_\mathrm{r}$ (red) for $\gamma\!=\!2.8$. Filled circles denote experimental data, and solid lines are QMC calculations performed for the experimental conditions.
\textbf{c,} Correlation length $\xi$ (cyan) and modulation amplitude $A$ (magenta) extracted from fits of $G^{(1)}(x)$ as a function of lattice depth. The filled markers give experimental fit values, and solid curves the corresponding QMC fitting results with shaded bands indicating fitting uncertainties.
\textbf{d,} Visibility of the interference peaks as a function of lattice depth. The points are the experimental values, and the solid curve the corresponding QMC results. All experimental results are averaged over 20 realizations.
}
\end{figure*}

We next isolate the intrinsic coherence of the Mott state from the effects of finite temperature and harmonic confinement present in the experiment. For this, we perform path-integral QMC simulations for continuous space adding a 1D lattice Bose gas in a box trap at near-zero temperature, described by the many-body Hamiltonian
\begin{equation}
\hat{H}\!=\!\sum_{i\!=\!1}^{N}\left[-\frac{\hbar^{2}}{2m}\nabla_{i}^{2}+V_{x}\sin^{2}(\pi x_{i}/a)\right]
+g_{1\mathrm{D}}\sum_{i<j}\delta(x_{i}-x_{j}),
\label{eq1.Hamiltonian}
\end{equation}
where $g_{1\mathrm{D}}$ is the interaction coupling constant~\cite{dunjko2001bosons}.
At given particle density $n$, it is linked to $\gamma\!=\!m g_{1\mathrm{D}}/\hbar^2 n$.
In our calculation, we simulate the Mott regime at unit filling $na\!=\!1$. This clean-limit calculation gives distributions $n(k)$ that agree well with the measured ones for different values of the lattice depth, see inset of Fig.~\ref{fig:1}b. Notably, at $V_x\!=\!5E_\mathrm{r}$, where the simulated system is a homogeneous MI phase with zero superfluid fraction, the distribution $n(k)$ exhibits pronounced interference peaks. Their appearance in a nearly pure Mott state indicates that the interference is an intrinsic property of the MI.

To establish that the observed interference peaks indeed arise within a gapped insulator, we record the excitation spectrum using the lattice-modulation technique~\cite{greiner2002,haller2010pinning,stoferle2004transition}. At $V_x\!=\!1.5E_\mathrm{r}$, we use a $20\%$ sinusoidal modulation of $V_x$ and scan the modulation frequency $f$ while monitoring the resulting heating through a broadening of $n(k)$. As shown in Fig~\ref{fig:2}a, for weak interactions $\gamma\approx0.5$ ($a_s\!=\!60a_0$), the heating response is weak and nearly linear with $f$, consistent with the presence of a gapless superfluid. In contrast, at $\gamma\!=\!2.8$ ($a_s\!=\!220a_0$), the heating response remains weak below a threshold frequency, but rises rapidly once this threshold is crossed. This threshold behavior signals the opening of an excitation gap, showing that the pinning transition has already occurred well below $1.5E_\mathrm{r}$ and that the states probed at $V_x\!=\!5E_\mathrm{r}$ lie deeply in the Mott-insulating regime.


We next examine how the interference pattern depends on the lattice depth. Experimentally, as $V_x$ is increased from $0E_\mathrm{r}$ to $5E_\mathrm{r}$ at fixed strong interactions, the interference peaks become progressively more pronounced, as shown in Fig.~\ref{fig:2}b. Strikingly, this enhancement occurs as the system is driven deeper into the insulating regime and acquires a higher MI fraction, see Fig.~\ref{fig:2}c. To quantify the strengthening of the interference, we define the side-peak visibility $\mathcal{V}\!=\!-k_L^2\left.(d^2 n(k)/dk^2)\right|_{k\!=\!\pm 2k_L}/n(0)$ from the local curvature at the $\pm2k_L$-position, and find experimentally that $\mathcal{V}$ increases with $V_x$. Our QMC simulations, which take the harmonic confinement into account, reproduce the same trend, identifying the Mott phase itself rather than residual superfluidity as the origin of the interference.

The central physical picture is illustrated in Fig.~\ref{fig:4}a. In the conventional picture of the SF-MI transition, the superfluid is usually viewed as a coherent matter wave extending across the lattice, whereas the Mott insulator is reduced to an array of localized particles pinned to individual sites. Our observations, however, show that this dichotomy is incomplete. Each lattice site $j$ contains a phase factor $\theta_j$ that follows a certain order and cannot be neglected, although the order is short-ranged and exhibits exponential decay. Such a wave property encoded in the phase order $\theta_j$ is the core factor for the observed interference pattern of the insulator. 
Quantitatively, this can be understood by the one-body correlation function $g^{(1)}(x)\!=\!\langle \hat{\Psi}^\dagger(x)\hat{\Psi}(0)\rangle$.
In the standard low-energy description of 1D bosons, $g^{(1)}(x)$ can be viewed schematically as a product of density and phase contributions \cite{giamarchi2003quantum,RevModPhys.83.1405,10.21468/SciPostPhys.15.2.050},
\begin{equation}
g^{(1)}(x)\!\approx\! \sqrt{\langle \hat{\rho}^{\dagger}(x)\hat{\rho}(0)\rangle}\,
e^{\langle \hat{\theta}^{\dagger}(x)-\hat{\theta}(0)\rangle},
\label{eq2.g1}
\end{equation}
where $\hat{\rho}^{\dagger}(x)$ and $\hat{\theta}^{\dagger}(x)$ are local density and phase operators, respectively.
In a lattice, the density contribution $\langle \hat{\rho}^{\dagger}(x)\hat{\rho}(0)\rangle$ gives rise to a spatial modulation, while the phase term $e^{\langle \hat{\theta}^{\dagger}(x)-\hat{\theta}(0)\rangle}$ governs the decay of the correlations~\cite{supplement}. In a conventional deep-lattice MI, this decay is too rapid for the oscillatory structure to become visible. In the shallow-lattice pinning Mott regime realized here, however, the decay is sufficiently slow so that the oscillatory component of $g^{(1)}(x)$ survives and generates the side peaks in momentum space.

Guided by this picture, we reconstruct the one-body correlation function by Fourier transforming the measured $n(k)$. Although the experiment is performed in a harmonic trap where translational invariance is lost, the measured integrated correlation function, $G^{(1)}(x)\!=\!\int \langle \hat{\Psi}^{\dagger}(x+x_0)\hat{\Psi}(x_0)\rangle dx_0$, still retains the same underlying physics \cite{guo2024observation,doi:10.1126/sciadv.adk6870}. The measured $G^{(1)}(x)$, shown in Fig.~\ref{fig:4}b over a relevant range of lattice depths, exhibits the two features: an overall exponential decay together with a pronounced modulation at the lattice period. As $V_x$ is increased, the decay becomes steeper, while the oscillatory modulation remains clearly visible. To benchmark these measurements, we compare them with QMC simulations assuming experimental conditions~\cite{supplement}. The temperature is fixed by matching the lattice-free $G^{(1)}(x)$ of the 1D gas \cite{doi:10.1126/sciadv.adk6870}, yielding $T\!=\!10\,\mathrm{nK}$, and is then kept fixed for all lattice depths, so that the subsequent calculations contain no additional free parameters. As shown in Fig.~\ref{fig:4}b, the agreement between experiment and theory is excellent throughout.


Fitting both the experimental and simulated results with the form $f(x)\!=\!(1+A\sin(2\pi x/a))e^{-x/(a\xi)}$, with $\xi$ the correlation length in units of $a$ and $A$ the modulation amplitude, allows us to quantify these two ingredients encoded in $G^{(1)}(x)$. These parameters extracted from experiment and theory agree well over the full range of $V_x$. As the lattice is deepened, the $\xi$ decreases monotonically, while the modulation amplitude $A$ grows steadily, as shown in Fig.~\ref{fig:4}c. The one-body correlations therefore become shorter-ranged but more strongly modulated at the lattice scale. This identifies the oscillatory real-space structure of the Mott state as the microscopic origin of the side peaks in momentum space, consistent with the subleading harmonic structure in 1D field theory. This interplay between shrinking correlation length and growing modulation amplitude also explains the evolution of the visibility $\mathcal{V}$ of interference in momentum-space. The $\mathcal{V}$ first rises rapidly upon entering the Mott regime, reaches a maximum around $5E_\mathrm{r}$, and then decreases again, as shown in Fig.~\ref{fig:4}d. The interference is therefore governed not by a single order parameter, but by the competition between these two ingredients. Even when $n(k)$ becomes nearly Gaussian in the deep-lattice limit (e.g. Fig.\ref{fig:1}b), $G^{(1)}(x)$ can still retain an oscillatory short-range structure. Evidently, the Mott state itself maintains a directly observable wave character.

In conclusion, our results show that a Mott insulator is not merely the particle-like counterpart of a superfluid, but can retain a directly observable quantum wave character. In the strongly interacting 1D pinning regime, this wave character manifests itself as interference in momentum space and as oscillatory short-range correlations in real space. By identifying the real-space origin of this interference, our results provide experimental access to the subleading harmonic structure predicted in Haldane’s description of 1D bosons, and show that the pinning Mott regime provides an especially favorable setting in which density ordering enhances its visibility without completely erasing the underlying phase structure. Our work reveals that the disappearance of transport does not imply the loss of phase structure, and that insulating quantum matter can preserve visible interference patterns beyond the conventional picture. These results open a route toward studying how such hidden phase structure evolves far from equilibrium, including after interaction or lattice quenches~\cite{villa2019,bernier2014,cheneau2012light}
, and how it survives in more complex insulating regimes such as disordered Bose glasses~\cite{giamarchi1988,yao-boseglass-2020,yu2024observing} or dimensional crossover systems \cite{guo2024observation,chauveau2023superfluid,tao2023}.

\begin{acknowledgments}
The Innsbruck team acknowledges funding by the European Research Council (ERC) under project number 10120161 and by an FFG infrastructure grant with project number FO999896041. Y.G. is supported by the Austrian Science Fund (FWF) with project number 10.55776/COE1, and Quantum Science and Technology-National Science and Technology Major Project of China (Grant No.2025ZD0300400). H. Y. is supported by The Fundamental Research Funds for the Central Universities, Peking University. Z.W. is supported by the Quantum Science and Technology-National Science and Technology Major Project (Grant No. 2021ZD0302003) and the National Natural Science Foundation of China (Grant Nos. 12488301, 12034011, U23A6004). T. G. acknowledges funding from the Swiss National Science Foundation under grant number 200020-219400. Y.Z. is supported by the Austrian Academy of Sciences (\"OAW) with APART-MINT 12234.

\textbf{Author Contributions:} This work was conceived by Y.G., H.Y., H.C.N., T.G. M.L. and X.Y. Experiments were prepared by X.Y., S.D., M.H. Y.Z. and performed by X.Y. and Y.G. Data were analyzed by Y.G., X.Y., Z.W., W.C. and I.Z. Numerical simulations and analytical prediction were performed by C.W. and H.Y. The manuscript was drafted mainly by Y.G., H.Y. X.Y., C.W. and H.C.N. All authors contributed to the discussion and finalization of the manuscript. 
{\bf Data Availability:} The data shown in the main text are available via Zenodo~\cite{zenodo_data_availability}.
{\bf Code Availability:} 
The Quantum Monte Carlo algorithm makes use of the Algorithms and Libraries for Physics Simulations (ALPS) scheduler library and statistical analysis tools \cite{troyer1998parallel,albuquerque2007alps,bauer2011alps}.
Codes supporting the findings of this study are available from the corresponding author upon reasonable request.

\emph{Note}---As we were finalizing our paper, we found that R. Vatré {\it et al.}~\cite{vatre2026phase} had independently posted a preprint on the similar topic.

\end{acknowledgments}
\bibliography{apssamp}

\begin{thebibliography}{47}
\bibitem [{zen()}]{zenodo_data_availability}%
  \BibitemOpen
  \href@noop {} {}\bibinfo {note} {Data set is available from Zenodo at 10.5281/zenodo.20067233}\BibitemShut {NoStop}%
\end{thebibliography}

\newpage
\clearpage
\newpage
\appendix
\renewcommand{\appendixname}{}
\section*{\centering Supplemental Materials}
\setcounter{figure}{0}         
\renewcommand{\thefigure}{S.\arabic{figure}}


\section{Experimental Parameters}

The BEC is produced in an optical dipole trap with trapping frequencies $\omega_{x,y,z}\!=\!(11.3(6),\,10.0(5),\,14.2(8))\times2\pi, \mathrm{Hz}$. To prepare a near-unit-filling initial state, the gas is loaded into a 3D optical lattice using a 1.5-s exponential ramp at $a_s=300\,a_0$, followed by a 100-ms linear magnetic-field ramp to the final interaction strength. The crossover to the 1D geometry is then achieved by exponentially lowering $V_x$ over 200 ms, while keeping the transverse lattices deep, resulting in an array of independent tubes with a weighted average of $\sim 33$ atoms per tube. Before detection, the system is held for 20 ms to equilibrate, after which all lattice potentials are switched off and the scattering length is quenched to zero. The atoms are then probed by 50-ms ToF absorption imaging. To quantify the energy gap of MI in the shallow-lattice, we apply the standard technique of lattice-modulation spectroscopy \cite{greiner2002,haller2010pinning}. The longitudinal lattice depth $V_x=1.5 E_{\rm r}$ is modulated sinusoidally for 50 ms with an amplitude of $0.3 E_{\rm r}$. After modulation, $n(k)$ is obtained by standard ToF imaging, and its width is extracted from Lorentzian fits. By comparing the broadened width to the unmodulated reference, we obtain the excitation spectrum. In the Mott-insulating regime, the finite excitation gap suppresses the response below a threshold frequency, resulting in a distinct threshold in the width vs. frequency response, a robust signature of the MI phase.


\section{QMC simulation}
The numerical results for the one-body correlation function and density profile are obtained directly via the Path Integral Monte Carlo (PIMC) method. The distribution $n(k)$ is computed by performing a Fourier transform on the one-body correlation function $g^{(1)}(x)$.  We conduct simulations of the single-tube Hamiltonian (Eq.~\ref{eq1.Hamiltonian}) at temperature $T$, 1D interaction strength $g_{\rm 1D}$, and chemical potential $\mu$, corresponding to the particle number $N$, within the grand-canonical ensemble. The interaction propagator is evaluated using the pair-product approximation, which is exact for delta-function interactions \cite{PhysRevA.91.043607}. For the ideal condition result shown in Fig.~\ref*{fig:1}, we set $T\!\approx\!0.5$ nK, which is near-zero temperature, and the system is at unit filling. In Fig.~\ref*{fig:2}, while keeping the temperature near zero, we take into account the harmonic trap present in the experiment and fix the particle number at  $N\!=\!50$, reflecting the situation of the center tube in the experiment setup. Finally, for comparing the measured $G^{(1)}(x)$ at different $V_x$ shown in Fig.~\ref*{fig:4}, we use the full experimental parameters, including the harmonic trap, the finite temperature of $10$nK, which is determined by the thermometry detailed in Ref.~\cite{doi:10.1126/sciadv.adk6870} and an averaged particle number $N\!=\!33$ across all the tubes. 

\section{Haldane's harmonic-fluid approach for correlation function}

The low-energy properties of 1D interacting bosons can be universally described using the harmonic fluid approach, where the system is characterized by two collective fields: the density $\hat{\rho}(x)$ and the phase $\hat{\theta}(x)$. These fields satisfy the canonical commutation relation $[\hat{\rho}(x), \hat{\theta}(x')] = i\delta(x-x')$. In this framework, the bosonic field operator $\hat{\Psi}^{\dagger}(x)$ is expressed in terms of these fields as an infinite series of harmonics \cite{haldane1981effective,RevModPhys.83.1405} 
\begin{equation}
\hat{\Psi}^{\dagger}(x) \simeq \left[ \rho_{0}-\frac{1}{\pi} \partial_{x} \hat{\phi}(x) \right]^{1 / 2} \sum_{m} \beta_{m} e^{2 i m(\pi \rho_{0} x-\hat{\phi}(x))} e^{-i \hat{\theta}(x)},
\end{equation}
where $\rho_0$ is the average density, $\hat{\phi}(x)$ and $\hat{\theta}(x)$ are the dual bosonic fields, and $\beta_m$ are non-universal coefficients. This representation accounts for the discrete nature of the particles through the summation over $m$. By utilizing this field-operator expansion, the $g_1(x) = \langle \hat{\Psi}^{\dagger}(x) \hat{\Psi}(0) \rangle$ for a finite system of length $L$ with periodic boundary conditions can be analytically derived. Its asymptotic behavior is given by \cite{haldane1981effective,RevModPhys.83.1405}:
\begin{equation}
\begin{aligned}
g_{1}(x) = &\rho_{0} \left[ \frac{1}{\rho_{0} d(x|L)} \right]^{\frac{1}{2K}} \left\{ A_{0} + \sum_{m=1}^{\infty} A_{m} \right. \\
&\times \left. \left[ \frac{1}{\rho_{0} d(x|L)} \right]^{2m^2 K} \cos(2\pi m \rho_{0} x) \right\}
\end{aligned}
\end{equation}
where $d(x|L) = \frac{L}{\pi} |\sin(\frac{\pi x}{L})|$ is the chord function for periodic boundary conditions, $K$ is the Luttinger parameter, and $A_m$ are non-universal constants. Beyond the leading term, the subleading harmonic contributions generate oscillatory structure in $g^{(1)}(x)$. In our system, these oscillations provide a natural real-space origin for the interference patterns observed in the $n(k)$, even in the Mott-insulating regime.

\begin{figure}[ht!]
\includegraphics[width=0.49\textwidth]{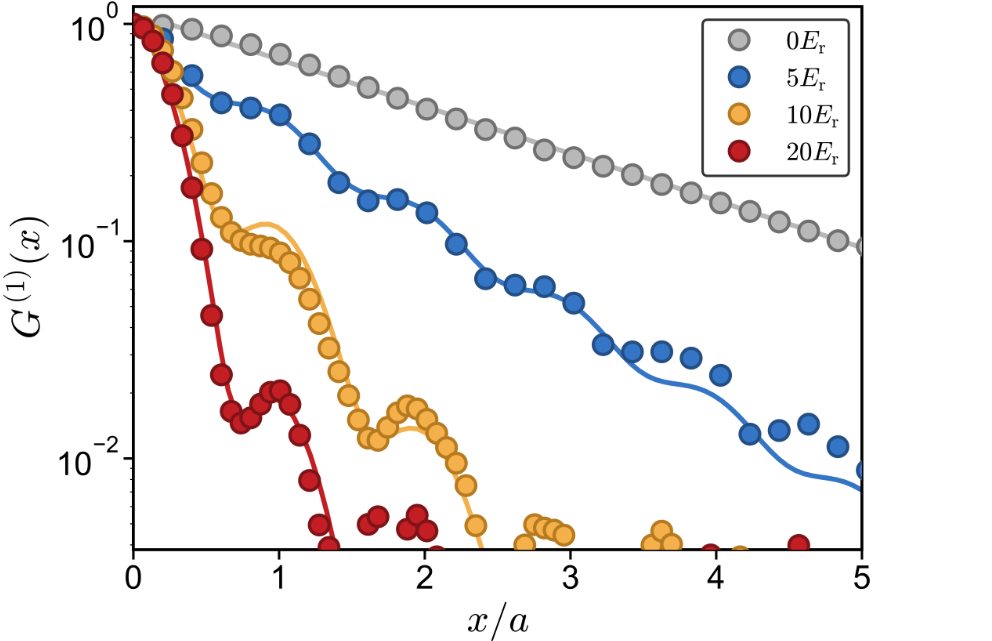}
\caption{\label{fig:fitting} \textbf{Fits to the one-body correlation function.}
Measured $G^{(1)}(x)$ as a function of distance $x/a$ for lattice depths $0E_\mathrm{r}$ (grey), $5E_\mathrm{r}$ (blue), $10E_\mathrm{r}$ (gold), and $20E_\mathrm{r}$ (red). Circles denote experimental data and solid lines the corresponding fits.}
\end{figure}

\section{Fitting of the one-body correlation function}

To extract the correlation length $\xi$ and modulation amplitude $A$ from the experimental data, we perform least-squares fits in logarithmic space. Because $G^{(1)}(x)$ decays exponentially over several orders of magnitude, a direct fit in linear would be dominated by the large values at short distance and would underweight the oscillatory structure at larger $x/a$. We therefore minimize the residuals between $\ln[G^{(1)}(x)]$ and $\ln[f(x)]$, where $f(x) = [1+A\sin(2\pi x/a)]e^{-x/(a\xi)}$ is mentioned in the main text.
This procedure balances the contribution from the short-distance oscillation and the long-distance tail, and thus allows both the exponential envelope and the lattice-periodic modulation to be determined reliably. As shown in Fig.~\ref{fig:fitting}, the resulting fits accurately follow the experimental data over the full dynamic range, from the central peak to the noise floor at the $10^{-3}$ level.

\end{document}